\def\rfr#1{eq. (\ref{#1})}

\def\Rfr#1{Eq. (\ref{#1})}

\def\dert#1#2{\frac{{{d}}{#1}}{{{d}}{#2}}}              

\def\bar{\begin{eqnarray}}
\def\ear{\end{eqnarray}}
\def\bb{\bibitem}
\def\eqi{\begin{equation}}
\def\eqf{\end{equation}}
\def\eqia{\begin{eqnarray}}
\def\eqfa{\end{eqnarray}}
\def\rp#1#2{{#1\over#2}}

\def\lb#1{\label{#1}}

\def\oc2{$\mathcal{O}(c^{-2})$}

\documentclass[11pt]{article}
\usepackage{amsmath,amsthm,amscd,amssymb}
\usepackage{latexsym}
\usepackage{graphicx,epsfig}

\begin{document}

\noindent{\bf \LARGE{Can the Pioneer anomaly be of gravitational
origin? A phenomenological answer}}
\\
\\
\\
{Lorenzo Iorio}\\
{\it Viale Unit$\grave{a}$ di Italia 68, 70125\\Bari (BA), Italy
\\tel. 0039 328 6128815
\\e-mail: lorenzo.iorio@libero.it}

\begin{abstract}
In order to satisfy the equivalence principle, any
non-conventional mechanism proposed to gravitationally explain the
Pioneer anomaly, in the form in which it is presently known from
the so-far analyzed Pioneer 10/11 data, cannot leave out of
consideration its impact on the motion of the planets of the Solar
System as well, especially those orbiting in the regions in which
the anomalous behavior of the Pioneer probes manifested itself. In
this paper we, first, discuss the residuals of the right ascension
$\alpha$ and declination $\delta$ of Uranus, Neptune and Pluto
obtained by processing various data sets with different, well
established dynamical theories (JPL DE, IAA EPM, VSOP). Second, we
use the latest determinations of the perihelion secular advances
of some planets in order to put on the test two gravitational
mechanisms recently proposed to accommodate the Pioneer anomaly
based on two models of modified gravity. Finally, we adopt the
ranging data to Voyager 2 when it encountered Uranus and Neptune
to perform a further, independent test of the hypothesis that a
Pioneer-like acceleration can also affect the motion of the outer
planets of the Solar System. The obtained answers are negative.
\end{abstract}



\section{Introduction}
\subsection{The Pioneer anomaly}
The so-called Pioneer anomaly (Anderson et al. 1998; 2002a)
consists of an unexpected, almost constant and uniform
acceleration  directed towards the Sun \eqi A_{\rm Pio}=(8.74\pm
1.33)\times 10^{-10} \ {\rm m\ s}^{-2}\lb{pioa}\eqf detected in
the so-far analyzed data of both the spacecraft Pioneer 10
(launched in March 1972) and Pioneer 11 (launched in April 1973)
after they passed the threshold of 20 Astronomical Units (AU; 1 AU
is slightly less than the average Earth-Sun distance and amounts
to about 150 millions kilometers), although it might also have
started to occur after 10 AU only, according to a recent analysis
of the Pioneer 11 data (Nieto and Anderson 2005). Latest
communications with the Pioneer spacecraft, confirming the
persistence of such an anomalous feature, occurred when they
reached 40 AU (Pioneer 11) and 70 AU (Pioneer 10). A thorough
re-analysis of all the available data sets of both the Pioneer
spacecraft (Turyshev et al. 2006a; 2006b) is currently ongoing and
should be completed within next year.
\subsection{Gravitational explanations of the Pioneer anomaly and planetary motions}
This anomalous effect recently attracted considerable attention
because of the possibility that it may be  a signal of some
failure in the currently accepted Newton-Einstein laws of
gravitation; indeed, at present no convincing explanations of it
in terms of either known gravitational effects or some
non-gravitational forces peculiar to the spacecraft themselves
were yet found.  A review of some of the proposed mechanisms of
gravitational origin see, e.g., (Anderson e al. 2002a; Dittus et
al. 2005). However, Murphy (1999) and Katz (1999) suggested
non-gravitational mechanisms which, in their intentions, would be
able to explain the Pioneer 10/11 anomalous behavior; see
(Anderson et al. 1999a; 1999b) for replies. Interesting
considerations about the energy transfer process in planetary
flybys and their connection with the Pioneer anomaly can be found
in (Anderson et al. 2007).

If the Pioneer anomaly is of gravitational origin, it must, then,
fulfil the equivalence principle, which is presently tested at a
$10^{-12}$ level (Will 2006) and lies at the foundations of the
currently accepted metric theories of gravity. In its weak form,
it states that different bodies fall  with the same acceleration
in a given external gravitational field. As a consequence,
whatever gravitational mechanism is proposed to explain the
investigated effect, it must also act, in general, on the Solar
System bodies and, in particular, on those planets whose orbits
reside in the region in which the Pioneer anomaly manifested
itself, according to what we presently know about it.

In this framework, Jaekel and Reynaud (2005a; 2005b) put forth an
ingenious gravitational mechanism able to accommodate the Pioneer
anomaly; it is based on a suitable metric linear extension of
general relativity which yields, among other things, an
acceleration only affecting the radial component of the velocity
of a test particle. In (Jaekel and Reynaud 2006a) a further,
non-linear extension of such a model was proposed and used. The
last attempt by Jaekel and Reynaud (2006b) to find a
non-conventional explanation of gravitational-type of the Pioneer
anomaly was based on an extra-potential  quadratic in distance.
Brownstein and Moffat (2006), instead, adopted a Yukawa-like,
explicit analytical model for an extra-acceleration acting upon a
test particle involving four free parameters, and fitted it to all
the presently available Pioneer 10/11 data points.

In this paper we will focus on such proposals by explicitly
deriving theoretical predictions of some dynamical orbital
effects, and by comparing them with the latest determinations
(Section \ref{JR} and Section \ref{BM}). The Russian astronomer
E.V. Pitjeva (Institute of Applied Astronomy, Russian Academy of
Sciences) recently processed almost one century of data of all
types in the effort of continuously improving the EPM2004
planetary ephemerides (Pitjeva 2005a). Among other things, she
also determined anomalous secular, i.e. averaged over one orbital
revolution, advances of the perihelia $\Delta\dot\varpi_{\rm det}$
of the inner (Pitjeva 2005b) and of some of the outer (Pitjeva
2006a; 2006b) planets as fit-for parameters\footnote{The
perihelia, as the other Keplerian orbital elements, are not
directly observable. } of global solutions in which she
contrasted, in a least-square way, the observations (ranges,
range-rates, angles like right ascension $\alpha$ and declination
$\delta$ , etc.) to their predicted values computed with a
complete suite of dynamical force models including all the known
Newtonian and Einsteinian features of motion\footnote{Only the
general relativistic, gravitomegnetic Lense-Thirring effect and
the Newtonian force due to the Kuiper belt objects (in the case of
the inner planets) were not modelled. In regard to the other
dynamical accelerations, the general relativistic gravitoelectric
field and the Newtonian effect due to the Sun's oblateness were
included by keeping the PPN parameters $\beta$ and $\gamma$ fixed
to 1 and $J_2=2\times 10^{-7}$, respectively.}. Thus, any
unmodelled force, as it would be the case for a Pioneer-like one
if present in Nature, is entirely accounted for by the determined
perihelia extra-advances. In regard to the outer planets, Pitjeva
was able to determine the extra-advances of perihelion for
Jupiter, Saturn and Uranus (see Table \ref{tavolapar} for their
relevant orbital parameters)
{\small\begin{table}\caption{ Semimajor axes $a$, in AU,
eccentricities $e$ and orbital periods $P$, in years, of the outer
planets of the Solar System.  Modern data sets covering at least
one full orbital revolution currently exist only for Jupiter,
Saturn and Uranus. }\label{tavolapar}

\begin{tabular}{llllll} \noalign{\hrule height 1.5pt}

 & Jupiter & Saturn  & Uranus & Neptune & Pluto\\
$a$ & 5.2 & 9.5 & 19.19 & 30.06 & 39.48\\
$e$ & 0.048 & 0.056 & 0.047 & 0.008 & 0.248\\
$P$ & 11.86 & 29.45 & 84.07 & 163.72 & 248.02\\
\hline

\noalign{\hrule height 1.5pt}
\end{tabular}

\end{table}}
because the temporal extension of the used data set covered at
least one full orbital revolution just for such planets: indeed,
the orbital periods of Neptune and Pluto amount to about 164  and
248 years, respectively. For the external regions of the Solar
System only optical observations were used, apart from Jupiter
(Pitjeva 2005a); they are, undoubtedly, of poorer accuracy with
respect to those used for the inner planets which also benefit of
radar-ranging measurements, but we will show that they are
accurate enough for our purposes. In regard to Uranus and Neptune,
in Section \ref{voyager} we will use  certain short-period, i.e.
not averaged over one revolution, effects of their semimajor axes
$a$ and  the ranging distance measurements to them performed at
Jet Propulsion Laboratory (JPL), NASA, during their encounter with
the Voyager 2 spacecraft (Anderson et al. 1995).
\subsection{Previous planetary data analysis and the Pioneer anomaly}
The idea of looking at the impact of a Pioneer-like acceleration
on the orbital dynamics of the Solar System bodies was put forth
for the first time by Anderson et al. (1998; 2002a) who, however,
considered the motion of the Earth and Mars finding no evidence of
any effect induced by an extra-acceleration like that of
\rfr{pioa} (see also Section \ref{voyager}). More interestingly,
Anderson et al. (2002b) preliminarily investigated the effect of
an ever-present, uniform Pioneer-like force on the long-period
comets. Wright (2003) got another negative answer for Neptune from
an analysis of the ranging data of Voyager 2 (see Section
\ref{voyager}). Page et al. (2005) proposed to use comets and
asteroids to assess the gravitational field in the outer regions
of the Solar System and thereby investigate the Pioneer anomaly.
The first extensive analysis involving all the outer planets can
be found in (Iorio and Giudice 2006). In it the time-dependent
patterns of the true observable quantities $\alpha\cos\delta$ and
$\delta$ induced by a Pioneer-like acceleration on Uranus, Neptune
and Pluto were compared with the observational residuals
determined in (Pitjeva 2005a) for the same quantities and the same
planets over a time span of about 90 years from 1913 (1914 for
Pluto) to 2003. While the former ones exhibited well defined
polynomial signatures yielding shifts of hundreds of arcseconds,
the latter ones did not show any particular patterns, being almost
uniform strips constrained well within $\pm 5$ arcseconds over the
data set time span which includes the entire Pioneer 10/11
lifetimes. An analogous conclusion can also be found in (Tangen
2006), although a different theoretical quantity was used in the
comparison with the data. It should also be remarked that the very
same conclusion could already have been obtained long time ago  by
using the residuals of some sets of modern optical observations
(1984-1997) to the outer planets processed by Morrison and Evans
(1998) with the  NASA JPL DE405  ephemerides (Standish 1998):
indeed, the residuals of $\alpha\cos \delta$ and $\delta$, shown
in Figure 4 of (Morrison and Evans 1998), are well within $\pm
0.4$ arcseconds. Analysis of residuals obtained with even older
ephemerides would have yielded the same results. For example,
Standish (1993) used JPL DE200 ephemerides (Standish 1982) to
process optical data of Uranus dating back to 1800: the obtained
residuals of $\alpha$ and $\delta$ do not show any particular
structure being well constrained within $\pm 5$ arcseconds. Gomes
and Ferraz-Mello (1987) used the VSOP82 ephemerides (Bretagnon
1982) to process more than one century (1846-1982) of optical data
of Neptune getting no anomalous signatures as large as predicted
by the presence of a Pioneer-like anomalous force. In regard to
Pluto, Gemmo and Barbieri (1994) and Rylkov et al. (1995) used the
JPL DE200 and JPL DE202  ephemerides (Standish 1990) in producing
residuals of $\alpha$ and $\delta$: no Pioneer-type signatures can
be detected in them.
\section{The Jaekel and Reynaud models}\lb{JR}
\subsection{The linear model}
In order to find an explanation of gravitational origin for the
anomalous acceleration  experienced by the Pioneer 10/11
spacecraft, Jaekel and Reynaud (2005a; 2005b) proposed, as a first
attempt, to use a suitable metric linear extension of general
relativity with two potentials $\Phi_{\rm N}$ and $\Phi_{\rm P}$.
In the gauge convention of the PPN formalism its space-time line
element, written in isotropic spherical coordinates, is (Jaekel
and Reynaud 2005b) \eqi
ds^2=g_{00}c^2dt^2+g_{rr}[dr^2+r^2(d\theta^2+\sin^2\theta
d\phi^2)],\eqf with
\begin{equation}\left\{\begin{array}{lll}g_{00}=1+2\Phi_{\rm
N},\\\\
g_{rr}=-1+2\Phi_{\rm N}-2\Phi_{\rm
P}.\lb{poti}\end{array}\right.\end{equation} In order to
accommodate the Pioneer anomaly,  the following simple model \eqi
\Phi_j(r)=-\rp{G_j M }{c^2 r}+\rp{\zeta_j M r}{c^2}, j={\rm N,
P},\lb{potii}\eqf was used (Jaekel and Reynaud 2005a; 2005b). It
is determined by four constants: the Newtonian constant $G_{\rm
N}$ and the three small parameters $G_{\rm P}, \zeta_{\rm N }$ and
$\zeta_{\rm P}$ which measure the deviation from general
relativity. In the intentions of Jaekel and Reynaud, their theory
should be able to explain the occurrence of the Pioneer anomaly
 a) without violating either the existing constraints from the
planetary motions b) or the equivalence principle. The latter goal
was ensured by the metric character of the proposed extension of
general relativity. In regard to a), they, first, focussed their
attention to the modification of the Newtonian potential. By using
the orbits of Mars and the Earth they got an upper bound $|\zeta
_{\rm N}M|\simeq 5\times 10^{-13}$ m s$^{-2}$ (Jaekel and Reynaud
2005b) which excludes that $\zeta_{\rm N}M r/c^2$ is capable to
account for the anomalous Pioneer acceleration. The key point of
their line of reasoning in explaining the Pioneer anomaly without
contradicting our knowledge of the planetary orbits consisted in
considering from the simple expression of \rfr{potii} for
$\Phi_{\rm P}$ the following extra-kinetic radial
acceleration\footnote{The contribution of $G_{\rm P}$ was found to
be negligible.} \eqi A_{\rm JR }=2\zeta_{\rm P}M\rp{v^2_r}{c^2},
\lb{accelk}\eqf where $v_r$ is the radial component of the
velocity of the moving body, in identifying it with the source of
the Pioneer anomalous acceleration by getting\footnote{The almost
constant value $v_r=1.2\times 10^4$ m s$^{-1}$ was used for both
the Pioneer spacecraft.} $\zeta_{\rm P}M=0.25$ m s$^{-2}$ and in
claiming that \rfr{accelk} cannot affect the planetary motions
because circular. Conscious of the fact that independent tests are
required to support their hypothesis and since no accurate and
reliable data from other spacecraft are available to this aim,
Jaekel and Reynaud (2005b) proposed to perform light deflection
measurements because $\Phi_{\rm P}$ affects the motion of
electromagnetic waves as well. A re-analysis of the Cassini
(Bertotti et al. 2003) data was suggested (Jaekel and Reynaud
2005b).

In this Section we will show that, in fact, the extra-kinetic
acceleration of \rfr{accelk} does also affect the orbital motions
of the planets in such a way that it is possible to compare the
resulting features of motion with the latest data from planetary
ephemerides, thus performing right now a clean and independent
test of the hypothesis that \rfr{accelk} is able to accommodate
the Pioneer anomaly. We will also discuss the feasibility of the
proposed light deflection measurements in view of the results
obtained from the perihelia test.
\subsubsection{The orbital effects of the kinetic acceleration and
comparison with the latest data}
In order to make a direct comparison with the extra-rates of
perihelia determined by Pitjeva (2005b), we will now analytically
work out the secular effects induced by the extra-kinetic
acceleration of \rfr{accelk} on the pericentre of a test particle.
To this aim, we will treat \rfr{accelk} as a small perturbation of
the Newtonian monopole. In order to justify this assumption, we
will, first, evaluate the average of \rfr{accelk} and, then, we
will compare it with the the Newtonian  mean accelerations
throughout the Solar System. To this aim, we must evaluate
\rfr{accelk} onto an unperturbed Keplerian ellipse by using \eqi
v_r=\rp{nae\sin f}{\sqrt{1-e^2}},\lb{vrad}\eqf where
$n=\sqrt{GM/a^3}$ is the (unperturbed) Keplerian mean motion and
$f$ is the true anomaly. Subsequently, the average over one
orbital period $P=2\pi/n$ has to  be performed. It is useful to
adopt the eccentric anomaly $E$ by means of the relations
\begin{equation}\left\{\begin{array}{lll}
dt=\rp{(1-e\cos E)}{n}dE
,\\\\
\cos f=\rp{\cos E-e}{1-e\cos E
},\\\\
\sin f=\rp{\sin E\sqrt{1-e^2}}{1-e\cos E}.
\lb{eccen}\end{array}\right.\end{equation}
With \eqi\int_0^{2\pi}\rp{\sin^2 E}{1-e\cos E }
dE=\rp{2\pi}{e^2}\left(1-{\sqrt{1-e^2}}\right),\eqf we get
\eqi\left\langle A_{\rm P}\right\rangle=\rp{2\zeta_{\rm P}M n^2
a^2}{c^2}\left(1-{\sqrt{1-e^2}}\right).\lb{aveap}\eqf \Rfr{aveap}
can, now, be compared with the averaged Newtonian monopole
acceleration\eqi \left\langle A_{\rm
N}\right\rangle=\rp{GM}{a^2\sqrt{1-e^2}},\lb{chepalle}\eqf The
results are in Table \ref{scassapalle} from which
{\small\begin{table}\caption{ Average Pioneer and Newtonian
accelerations for the Solar System planets, in m s$^{-2}$. For
$\left\langle A_{\rm P} \right\rangle$ the expression of
\rfr{aveap} was used with $\zeta_{\rm P}M=0.25$ m s$^{-2}$.
}\label{scassapalle}

\begin{tabular}{lll} \noalign{\hrule height 1.5pt}

Planet  & $\left\langle A_{\rm P} \right\rangle$ & $\left\langle A_{\rm N} \right\rangle$\\
Mercury & $2\times 10^{-10}$   & $4\times 10^{-2}$    \\
Venus   & $1\times 10^{-13}$   & $1\times 10^{-2}$    \\
Earth   & $6\times 10^{-13}$   & $6\times 10^{-3}$    \\
Mars    & $1\times 10^{-11}$   & $2\times 10^{-3}$    \\
Jupiter & $1\times 10^{-12}$   & $2\times 10^{-4}$    \\
Saturn  & $8\times 10^{-13}$   & $6\times 10^{-5}$    \\
Uranus  & $2\times 10^{-13}$   & $1\times 10^{-5}$    \\
Neptune & $6\times 10^{-15}$   & $6\times 10^{-6}$    \\
Pluto   & $4\times 10^{-12}$   & $4\times 10^{-6}$    \\
\hline

\noalign{\hrule height 1.5pt}
\end{tabular}

\end{table}}
it clearly turns out  that the use of the perturbative scheme is
quite adequate for our purposes. The Gauss equation for the
variation of $\varpi$ under the action of an entirely radial
perturbing acceleration $A_r$ is \eqi\dert\varpi t
=-\rp{\sqrt{1-e^2}}{nae}A_r\cos f.\lb{gaus}\eqf After being
evaluated onto the unperturbed Keplerian ellipse by using
\rfr{vrad}, \rfr{accelk} must be inserted into \rfr{gaus}; then,
the average over one orbital period has to be taken.
By means of \eqi \int_0^{2\pi}\rp{\sin^2 E(\cos E-e)}{(1-e\cos
E)^2}dE = \rp{2\pi}{e^3}\left(-2+e^2+2\sqrt{1-e^2}\right),\eqf it
is possible to obtain \eqi\left\langle\dert\varpi
t\right\rangle=-\rp{2\zeta_{\rm P}Mna\sqrt{1-e^2}}{c^2
e^2}\left(-2+e^2+2\sqrt{1-e^2}\right).\lb{pippoz}\eqf Note that
\rfr{pippoz} is an exact result, not based on approximations for
$e$. It may be interesting to note that the rates for the
semimajor axis and the eccentricity turn out to be zero; it is not
so for the mean anomaly $\mathcal{M}$, but no observational
determinations exist for its extra-rate. We will now use
\rfr{pippoz} and $\zeta_{\rm P}M=0.25$ m s$^{-2}$, which has been
derived from \rfr{accelk} by imposing that it is the source of the
anomalous Pioneer acceleration, to calculate the perihelion rates
of the inner
planets of the Solar System for which estimates of their
extra-advances accurate enough for our purposes exist (Pitjeva
2005b). The results are summarized in Table \ref{tavolazza}
{\small\begin{table}\caption{ (P): predicted extra-precessions  of
the longitudes of  perihelia of the inner planets, in arcseconds
per century, by using \rfr{pippoz} and $\zeta_{\rm P}M=0.25$ m
s$^{-2}$. (D): determined extra-precessions  of the longitudes of
perihelia of the inner planets, in arcseconds per century. Data
taken from Table 3 of (Pitjeva 2005b). It is important to note
that the quoted uncertainties are not the mere formal, statistical
errors but are realistic in the sense that they were obtained from
comparison of many different solutions with different sets of
parameters and observations (Pitjeva, private communication 2005).
}\label{tavolazza}

\begin{tabular}{lllll} \noalign{\hrule height 1.5pt}

 & Mercury & Venus  & Earth & Mars\\
(P) & 1.8323 & 0.001 & 0.0075 & 0.1906 \\
(D) & $-0.0036\pm 0.0050$ & $0.53\pm 0.30$ & $-0.0002\pm 0.0004$ & $0.0001\pm 0.0005$\\

\hline

\noalign{\hrule height 1.5pt}
\end{tabular}

\end{table}}

{\small\begin{table}\caption{ Values of $\zeta_{\rm P}M $, in m
s$^{-2}$, obtained from the determined extra-advances of perihelia
(Pitjeva 2005b). After discarding the value for Venus, the
weighted mean for the other planets yields $\zeta_{\rm P}M
=-0.0001\pm 0.0004$ m s$^{-2}$. The Pioneer anomalous acceleration
is, instead, reproduced for $\zeta_{\rm P}M=0.25$ m s$^{-2}$.
}\label{tavola2}

\begin{tabular}{lllll} \noalign{\hrule height 1.5pt}

 & Mercury & Venus  & Earth & Mars\\
$\zeta_{\rm P}M $ & $-0.0005\pm 0.0007$ & $91\pm 51$ & $-0.006\pm 0.013$ & $0.0001\pm 0.0006$\\
\hline

\noalign{\hrule height 1.5pt}
\end{tabular}

\end{table}}

It clearly turns out that the determined extra-advances of
perihelia are quite different from the values predicted in the
hypothesis that \rfr{accelk} can explain the Pioneer anomaly. In
Table \ref{tavola2} we show the values of $\zeta_{\rm P}M$ which
can be obtained from the determined extra-advances of perihelia
(Pitjeva 2005b); as can be noted, all of them are far from the
value which would be required to obtain the correct magnitude of
the anomalous Pioneer acceleration. The experimental intervals
obtained from Mercury, the Earth and Mars are compatible each
other; Venus, instead, yields values in disagreement with them.
This fact can be explained by noting that its perihelion is a bad
observable due to its low eccentricity ($e_{\rm Venus}=0.00677$).
By applying the Chauvenet criterion (Taylor 1997) we reject the
value obtained from the Venus perihelion since it lies at almost
2$\sigma$ from the mean value of the distribution of Table
\ref{tavola2}. The weighted mean for Mercury, the Earth and Mars
is, thus, $\left<\zeta_{\rm P}M\right>_{\rm w}=-0.0001$ m s$^{-2}$
with a variance, obtained from $1/\sigma^2=\sum_i(1/\sigma_i^2)$,
of 0.0004 m s$^{-2}$.

An analysis involving the perihelia of Mars only can be found in
(Jaekel and Reynaud 2006a). In it Jaekel and Reynaud presented a
nonlinear generalization of their model, and an explicit
approximate expression of the perihelion rate different from
\rfr{pippoz} can be found; it\footnote{More precisely, in (Jaekel
and Reynaud 2006a) an explicit expression for the adimensional
perihelion shift after one orbital period, in units of $2\pi$,
i.e. $(\dot\varpi P)/2\pi$, is present; a direct comparison with
our results can be simply done by multiplying their formula by $n$
and making the conversion from s$^{-1}$ to arcseconds per
century.} is calculated with $\zeta_{\rm P}M=0.25$ m s$^{-2}$
yielding a value for the Martian perihelion advance which is about
one half of  our value in Table \ref{tavola}. Even in this case,
the results by Pitjeva (2005b) for Mars would rule out the
hypothesis that the Pioneer anomaly can be explained by the
proposed nonlinear model. By the way, in (Jaekel and Reynaud
2006a) no explicit comparison with published or publicly available
data was presented.

All the previous considerations are based on the simple model of
\rfr{potii}, with $\zeta_{\rm P}$ constant over the whole range of
distances from the radius of the Sun to the size of the Solar
System. Jaekel and Reynaud (2005a; 2005b; 2006a), in fact, left
generically open the possibility that, instead, $\zeta_{\rm P}$
may vary with distance across the Solar System, but neither
specific empirical or theoretical justifications for such a
behavior were given nor any explicit functional dependence for
$\zeta_{\rm P}(r)$ was introduced. By the way, as already noted,
they explicitly applied their explicit model, with $\zeta_{\rm
P}M=0.25$ m s$^{-2}$, to the Mars' perihelion (Jaekel and Reynaud
2006a).
\subsubsection{The deflection of light} The results for $\zeta_{\rm
P}M$ from the determined extra-rates of the perihelia of the inner
planets allow us to safely examine the light deflection
measurements originally proposed by Jaekel and Reynaud as
independent tests of their theory; indeed, the values of Table
\ref{tavola2} certainly apply to the light grazing the Sun, also
in the case of an hypothetical variation of $\zeta_{\rm P}(r)$
with distance. Jaekel and Reynaud (2005a)  found the following
approximate expression for the deflection angle induced by
$\zeta_{\rm P}$ \eqi \psi_{\rm P}=-\rp{2\zeta_{\rm
P}M\rho}{c^2}L,\lb{def}\eqf where $\rho$ is the impact parameter
and $L$ is a factor of order of unity which depends
logarithmically on $\rho$ and on the distances of the emitter and
receiver to the Sun. For $\rho=R_{\odot}$, $L\sim 1$, and
$\zeta_{\rm P}M=-0.0001$ m s$^{-2}$, \rfr{def} yields a deflection
of only -0.3 microarcseconds, which can be translated into an
equivalent accuracy of about $2\times 10^{-7}$ in measuring the
PPN parameter $\gamma$ with the well-known first-order Einsteinian
effect (1.75 arcseconds at the Sun's limb). Such a small value is
beyond the presently available possibilities; indeed, the Cassini
test (Bertotti et al. 2003) reached a $10^{-5}$ level, which has
recently been questioned by Kopeikin et al. (2006) who suggest a
more realistic $10^{-4}$ error. Instead, it falls within the
expected 0.02 microarcseconds accuracy of the proposed LATOR
mission (Turyshev et al. 2006), which might be ready for launch in
2014. Also ASTROD (Ni 2002) and, perhaps, GAIA (Vecchiato et al.
2003), could reach the required sensitivity to measure such an
effect. However, because of technological and programmatic
difficulties, the launch of an ASTROD-like mission is not expected
before 2025. GAIA is scheduled to be launched in 2011
(http://gaia.esa.int/science-e/www/area/index.cfm?fareaid=26).
\subsection{The quadratic model and its confrontation with the
observationally determinations} 
\begin{figure}
\begin{center}
\includegraphics[width=14cm,height=11cm]{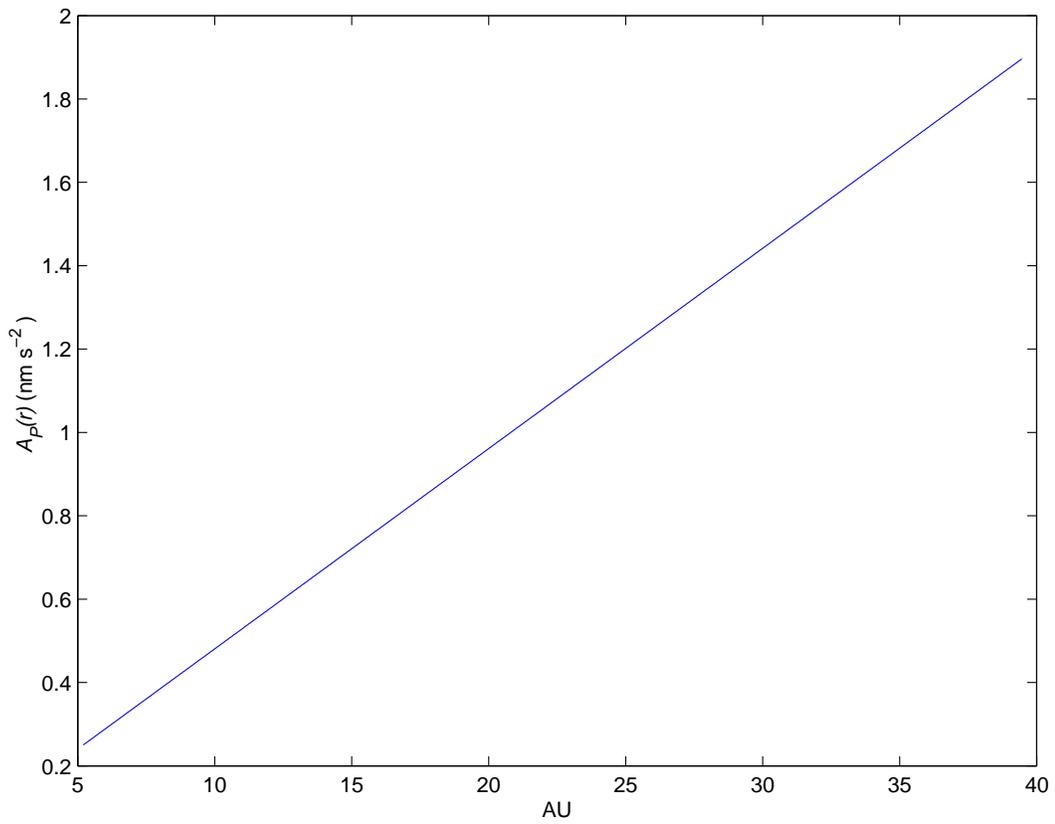}
\end{center}
\caption{\label{JRquad} Anomalous acceleration, in nm s$^{-2}$,
induced by $\delta\Phi_P=c^2\chi r^2$, with $\chi=4\times 10^{-8}$
AU$^{-2}$, according to Jaekel and Reynaud (2006b). }
\end{figure}

In (Jaekel and Reynaud 2006b) a
further development of the post-Einsteinian metric extension of
general relativity proposed by such authors is presented. It,
among other things, amends previous versions (Jaekel and Reynaud
2005a; 2005b) and yields another possible explanation of the
behavior experienced by the Pioneer probes. Indeed, it was found
that a roughly constant anomaly is produced when a second,
extra-potential $\delta\Phi_P(r)$, quadratic in $r$, is introduced
in the range of Pioneer distances. The choice by Jaekel and
Reynaud (2006b) was \eqi\delta\Phi_P(r)=c^2\chi r^2,\ \chi\simeq
4\times 10^{-8}\ {\rm AU}^{-2}, \eqf where  $c$ is the speed of
light in vacuum. The resulting acceleration \eqi A_P(r)=-2c^2\chi
r\lb{accnew},\eqf in units of nm s$^{-2}$, is plotted in Figure
\ref{JRquad}.
Without investigating how well such a model fits, in fact, all of
the currently available data of the Pioneer 10/11 data, here we
are going to derive theoretical predictions for the secular
perihelion advance induced by \rfr{accnew}. The standard methods
of perturbative celestial mechanics yield
\eqi\left\langle\dert\varpi t\right\rangle=-3c^2\chi\sqrt{
\rp{a^3(1-e^2)}{GM} }.\lb{perinew}\eqf  Note that \rfr{perinew} is
an exact result. The comparison among the anomalous advances for
Jupiter, Saturn and Uranus predicted with \rfr{perinew} and the
determined perihelia rates is in Table \ref{tavolazza2}.
{\small\begin{table}\caption{ First row: determined
extra-precessions of the longitudes of perihelia $\varpi$ of
Jupiter, Saturn and Uranus, in arcseconds per century (Pitjeva
2006a; 2006b). The quoted uncertainties are the formal,
statistical errors re-scaled by a factor 10 in order to get
realistic estimates. Second row: predicted anomalous
extra-precessions of the perihelia for Jupiter, Saturn and Uranus,
in arcseconds per century, according to \rfr{perinew}.
}\label{tavolazza2}

\begin{tabular}{llll} \noalign{\hrule height 1.5pt}

 & Jupiter & Saturn  & Uranus\\
(D) & $0.0062\pm 0.036$ & $-0.92\pm 2.9$ & $0.57\pm 13.0$\\
(P) & $-18.679$ & $-46.3$ & $-132.3$\\
\hline

\noalign{\hrule height 1.5pt}
\end{tabular}

\end{table}}

As can be noted, even by re-scaling by a factor 10 the formal
errors released by Pitjeva (2006a; 2006b), the discrepancy among
the predicted and the determined values amounts to 519, 15 and 10
sigma for Jupiter, Saturn and Uranus, respectively.

\section{The Brownstein and Moffat model}\lb{BM}
In order to explain the Pioneer anomaly, Brownstein and Moffat
(2006), in the context of their STVG metric theory of gravitation
(Moffat 2006a), considered a variation with distance of the
Newtonian gravitational constant $G(r)$ and proposed the following
radial extra-acceleration affecting the motion of a test particle
in the weak field of a central mass $M$ \eqi A_{\rm BM}=-\rp{G_0
M\xi(r)}{r^2}\left\{1-\exp\left[-\rp{r}{\lambda(r)}\right]\left[1+\rp{r}{\lambda(r)}\right]\right\}.\lb{accel}\eqf
Here $G_0$ is the `bare' value of the Newtonian gravitational
constant. Lacking at present a solution for $\xi(r)$ and
$\lambda(r)$, the following parameterization was introduced for
them\footnote{In the notation of Brownstein and Moffat (2006)
$\xi(r)$ and $d$ are $\alpha(r)$ and $\overline{r}$, respectively.
Note that there is an error in eq. (12) for $\lambda(r)$, p. 3430
of (Brownstein and Moffat 2006): a $-$ sign is lacking in front of
$b$. Instead, eq. (27) of (Moffat 2006b) gives the correct
expression.}
\begin{equation}\left\{\begin{array}{lll}\xi(r)=\xi_{\infty}\left[1-\exp\left(-\rp{r}{d}\right)\right]^{\rp{b}{2}},\\\\
\lambda(r)=\lambda_{\infty}\left[1-\exp\left(-\rp{r}{d}\right)\right]^{-b}.\lb{al}\end{array}\right.\end{equation}
In \rfr{al} $d$ is a scale distance and $b$ is a constant. The
best fitted values which reproduce the magnitude of the anomalous
Pioneer acceleration are (Brownstein and Moffat 2006)
\begin{equation}\left\{\begin{array}{lll}
\xi_{\infty}=(1.00\pm 0.02)\times 10^{-3},\\\\
\lambda_{\infty}=47\pm 1\ {\rm AU},\\\\
d=4.6\pm 0.2 \ {\rm AU},\\\\
b=4.0.
 \lb{fit}\end{array}\right.\end{equation}
The `renormalized' value $G_{\infty}$ of the Newtonian
gravitational constant-$G$ in the following-which is measured by
the usual astronomical techniques is related to the `bare'
constant by (Brownstein and Moffat 2006) \eqi
\rp{G_0}{G_{\infty}}=\rp{1}{ 1+\sqrt{\xi_{\infty}} }.\eqf With the
fit of \rfr{fit} we have \eqi \rp{G_0}{G_{\infty}}=0.96934.\eqf

The scope  of Brownstein and Moffat (2006) was to correctly
reproduce the Pioneer anomalous acceleration without contradicting
either the equivalence principle or our knowledge of the planetary
orbital motions. The first requirement was satisfied by the metric
character of their theory. In regard to the second point,
Brownstein and Moffat (2006) did not limit
  the validity of \rfr{accel} just to the region in which the Pioneer
anomaly manifested itself, but extended it to the entire Solar
System. Their model is not a mere more or less $ad\ hoc$ scheme
just to save the phenomena being, instead, rather `rigid' and
predictive. It is an important feature because it, thus, allows
for other tests independent of the Pioneer anomaly itself. This
general characteristic will also be  preserved in future, if and
when more points to be fitted will be obtained by further and
extensive re-analysis of the entire data set of the Pioneer
spacecraft (Turyshev et al. 2006a; 2006b) yielding a modification
of the fit of \rfr{fit}. Brownstein and Moffat (2006) performed a
test based on the observable
\eqi\eta=\left[\rp{G(a)}{G(a_{\oplus})}\right]^{1/3}-1,\eqf where
$a$ and $a_{\oplus}$ are the semimajor axes of a planet and of the
Earth. The quantity $\eta$ was related to the third Kepler's law
for which observational constraints existed from a previous
model-independent analysis (Talmadge et al. 1988) for the inner
planets and Jupiter. No observational limits were put beyond
Saturn because of the inaccuracy of the optical data used at the
time of the analysis by Talmadge et al. (1988). Brownstein and
Moffat (2006) found their predictions for $\eta$ in agreement with
the data of Talmadge et al. (1988).
\subsection{The confrontation with the observational determinations}
\begin{figure}
\begin{center}
\includegraphics[width=14cm,height=11cm]{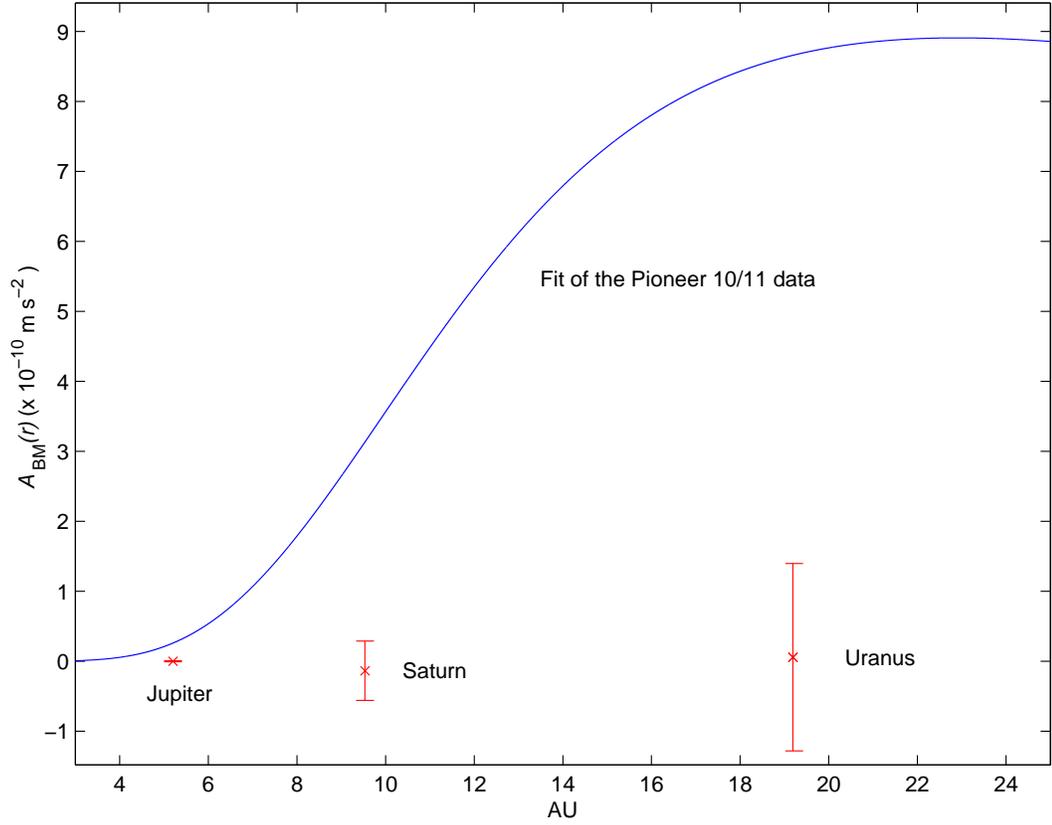}
\end{center}
\caption{\label{figura} The continuous curve is the fit to the
currently available Pioneer 10/11 data according to the model of
\rfr{accel} by Brownstein and Moffat (2006). The  anomalous
accelerations experienced by Jupiter, Saturn and Uranus, obtained
from the determined perihelion rates by Pitjeva (2006a; 2006b),
are also shown.}
\end{figure}

We will now perform an independent test of the Brownstein and
Moffat (2006) model by using the extra-rates of the perihelia of
Jupiter, Saturn and Uranus determined by Pitjeva (2006a; 2006b).
To this aim, it is important to note that the spatial variations
experienced by the extra-acceleration of \rfr{accel} over the
orbits of such planets are un-detectable because they amount to
just $0.1-0.01\times 10^{-10}$ m s$^{-2}$; thus, we will assume
the acceleration of \rfr{accel} to be uniform. From Table
\ref{scassapalle} it clearly turns out that the anomalous
acceleration of \rfr{accel} can be considered as a small
perturbation of the Newtonian monopole term which, indeed, is 6-4
orders of magnitude larger than it. In, e.g., (Iorio and Giudice
2006; Sanders 2006) it was shown that a radial, constant and
uniform perturbing acceleration $A_r$, induces a pericentre rate
\eqi \dert\varpi t=A_r\sqrt{\rp{a(1-e^2)}{GM}}.\lb{peri}\eqf We
will use \rfr{peri} and the determined extra-advances of
perihelion (Pitjeva 2006a; 2006b) in order to solve for $A$ and
compare the obtained values with those predicted by \rfr{accel}
for Jupiter, Saturn and Uranus. The results are summarized in
Table \ref{tavola} and Figure \ref{figura}
{\small\begin{table}\caption{ First row: determined
extra-precessions of the longitudes of perihelia $\varpi$ of
Jupiter, Saturn and Uranus, in arcseconds per century (Pitjeva
2006a; 2006b). The quoted uncertainties are the formal,
statistical errors re-scaled by a factor 10 in order to get
realistic estimates. Second row: predicted anomalous acceleration
for Jupiter, Saturn and Uranus, in units of $10^{-10}$ m s$^{-2}$,
according to the model of \rfr{accel} (Brownstein and Moffat
2006), evaluated at $r=a$. Third row: anomalous acceleration of
Jupiter, Saturn and Uranus, in units of $10^{-10}$ m s$^{-2}$,
from the determined perihelia precessions of the first row. The
quoted uncertainties have been obtained by means of the re-scaled
errors in the perihelia rates. Fourth row: discrepancy between the
determined and predicted accelerations in units of errors
$\sigma$. }\label{tavola}

\begin{tabular}{llll} \noalign{\hrule height 1.5pt}

 & Jupiter & Saturn  & Uranus\\
$\Delta\dot\varpi_{\rm det}$ & $0.0062\pm 0.036$ & $-0.92\pm 2.9$ & $0.57\pm 13.0$\\
$A_{\rm BM}(a)$ & 0.260 & 3.136 & 8.660\\
$A_{\rm det}$  & $0.001\pm 0.007$ & $-0.134\pm 0.423$ & $0.058\pm 1.338$ \\
$|A_{\rm det}-A_{\rm BM}(a)|/\sigma$& 37 & 7 &6\\

\hline

\noalign{\hrule height 1.5pt}
\end{tabular}

\end{table}}
As can be noted, the gravitational solution to the Pioneer anomaly
proposed by Brownstein and Moffat (2006), in the form of
\rfr{accel} and with the fitted values of \rfr{fit}, must be
rejected. Note that the quoted errors for the perihelia rates are
the formal uncertainties re-scaled by 10 in order to give
conservative evaluations of the realistic ones. In the case of
Jupiter even a re-scaling of 100 would still reject the value
predicted by \rfr{accel}.

Incidentally, let us note that the results by Iorio and Giudice
(2006) for Uranus, Neptune and Pluto can be viewed as a further,
negative test of the Brownstein and Moffat (2006) model because
they are based on the use of a Pioneer-like acceleration assumed
to be uniform in the regions crossed by such planetary orbits.

\section{A test with the Voyager 2 ranging data to Neptune}\lb{voyager}
Until now we used the secular perihelion advances of some of the
outer planets determined from optical data only (apart from
Jupiter). In this Section we will turn our attention to Neptune
and to certain short-period dynamical effects. The  ranging data
from Voyager 2 will be used as well (Anderson et al. 1995).

In (Iorio and Giudice 2006) there are the analytical expressions
of the short-period shifts induced on the Keplerian orbital
elements by a radial, constant perturbing acceleration $A_{r}$,
whatever its physical origin may be. For the semimajor
axis\footnote{Here and in the following we will assume $r\approx
a$, neglecting the finite value of the eccentricities, quite small
for Uranus and, especially, Neptune.} we have \eqi \rp{\Delta
a}{a}=-\rp{2eA_{r }a^2}{GM}(\cos E-\cos E_0
)=-\rp{2e}{\sqrt{1-e^2}}\rp{A_r}{\left\langle A_{\rm
N}\right\rangle}(\cos E-\cos E_0).\lb{pippo}\eqf In the following
computation it will be useful to express the eccentric anomaly $E$
in terms of the mean anomaly $\mathcal{M}$ as (Roy 2005)\eqi E
\sim \mathcal{M}+\left(e-\rp{e^3}{8}\right)\sin
\mathcal{M}+\rp{e^2}{2}\sin 2\mathcal{M}+\rp{3}{8}e^3\sin
3\mathcal{M}.\eqf The reference epoch is customarily assumed to be
J2000, i.e. JD=2451545.0 in Julian date. From \rfr{pippo} it can
be noted that, whatever the eccentricity of the orbit is,
  \eqi\left\langle\rp{\Delta a}{a}\right\rangle=0,\eqf so
  that $\Delta a/a$ cannot tell us anything about the impact of an acceleration
  like $A_{\rm Pio}$ for those planets for which data sets covering
  at least one full orbital revolution  exist. As already pointed out, to date, only Neptune
  and Pluto  have not yet described a full orbit since
  modern astronomical observations became available after the first decade of 1900.
Incidentally, let us note that,
according to \rfr{pippo}, $\Delta a/a=0$ for $e=0$.

  The situation is different for Neptune since no secular effects can yet be measured for it.
   Thus, let us use \rfr{pippo} and
  \rfr{pioa} for $A_r$
  getting
  \eqi \left.\rp{\Delta a}{a}\right|_{\rm Nep}=(-2.2882\pm 0.3482)\times 10^{-6}(\cos E-\cos E_0).\lb{effetto}\eqf
  Note that the anomalous acceleration predicted by  Brownstein and Moffat (2006)
   experiences an un-appreciable variation of just $0.04\times 10^{-10}$ m
  s$^{-2}$ over the Neptune's orbit, so that it can safely be
  considered uniform. Thus, the use of \rfr{effetto} can be considered as a test of the
  Brownstein and Moffat (2006) model as well, and of any other model capable of reproducing
  an extra-acceleration with the characteristics of
  \rfr{pioa} acting upon a test particle which moves in the spatial regions crossed  by the Neptune's orbit.
  The predicted effect of \rfr{effetto} can be compared with the
  latest available observational determinations. Pitjeva (2005a) used
  only
  optical data (Table 3 of (Pitjeva 2005a)) for the outer planets (apart from Jupiter) obtaining a
  formal, statistical error  $\delta a=478532$ m
  for the Neptune's semimajor axis (Table 4 of (Pitjeva 2005a)) at JD=2448000.5 epoch
  (Pitjeva 2006b).
  By re-scaling it by $10-30$
  times in order to get realistic uncertainty we
  get\eqi \left.\rp{\delta a}{a}\right|_{\rm Nep}^{(\rm optical)}=(1-3)\times
  10^{-6}.\eqf It must be compared with \rfr{effetto} at\footnote{For Neptune $E_0=128.571$ deg at JD=2451545.0.} JD=2448000.5 ($E=107.423$ deg)
\eqi\left.\rp{\Delta a}{a}\right|_{\rm Nep}({\rm
JD}=2448000.5)=(-0.7413\pm 0.1128)\times 10^{-6}.\eqf Such an
effect would be too small to be detected.

  In (Anderson et al. 1995) the ranging data of the Voyager 2 encounter with Neptune
  were
  used yielding a unique ranging measurement of $a$ (Julian Date
  JD=2447763.67); \rfr{effetto}, evaluated at such epoch ($E=106.012$ deg), predicts
  \eqi\left.\rp{\Delta a}{a}\right|_{\rm Nep}({\rm JD}=2447763.67)=(-0.7954\pm 0.1210)\times 10^{-6}.\lb{predi}\eqf
  By assuming for $\Delta a$ the residuals with respect to the DE200 JPL
  ephemerides used in Table 1 of (Anderson et al. 1995), i.e. $8224.0\pm 1$ km, one
  gets
  \eqi\left.\rp{\Delta a}{a}\right|_{\rm Nep}^{(\rm ranging)}=(1.8282\pm 0.0002)\times
  10^{-6}.\eqf
  This clearly rules out the prediction of \rfr{predi}.

  The same analysis can also be  repeated for Uranus ($P=84.07$ yr) for which no
  modern data covering a full orbital revolution were available at
  the time of the Anderson et al. (1995) work; as for Neptune, one ranging distance measurement is available
  from the Voyager 2 flyby with Uranus (JD=2446455.25). The prediction of
  \rfr{pippo}, with \rfr{pioa} for $A_r$, for the flyby epoch\footnote{For Uranus $E_0=70.587$ deg at JD=2451545.0.}
  ($E=8.860$ deg) is
\eqi\left.\rp{\Delta a}{a}\right|_{\rm Ura}({\rm
JD}=2446455.25)=(-3.3576\pm 0.5109)\times 10^{-6}.\lb{prediu}\eqf
Table 1 of (Anderson et al. 1995) yields for the DE200 residuals
of the Uranus' semimajor axis $\Delta a=147.3\pm 1$ km, so that
\eqi\left.\rp{\Delta a}{a}\right|^{(\rm ranging)}_{\rm Ura}({\rm
JD}=2446455.25)=(0.0513 \pm 0.0003)\times 10^{-6}.\lb{measu}\eqf
Also in this case, the effect which would be induced by $A_{\rm
Pio}$ on $\Delta a/a$ is absent. Since the variation of the
acceleration predicted by Brownstein and Moffat (2006) over the
Uranus orbit amounts to just $0.2\times 10^{-10}$ m s$^{-2}$, the
same considerations previously traced for Neptune hold for Uranus
as well.

It may be interesting to note that the paper by Anderson et al.
(1995) was used as a basis for other tests with the outer planets
using different methods. E.g., Wright (2003) and Sanders (2006)
adopted the third Kepler's law. Basically, the line of reasoning
is as follows. In the circular orbit limit, let us write, in
general, $P=2\pi a/v$; in particular, the third Kepler law states
that $P=2\pi\sqrt{a^3/K_p}$, where $K_p=GM_{\odot}$. If we assume
that $K_p$ may vary by $\Delta K_p$ for some reasons\footnote{E.g.
due to dark matter (Anderson et al. 1995).} inducing a change in
the orbital speed, then $\Delta v/v=(1/2)\Delta K_p/K_p$. In
general, for an additional radial acceleration acting upon a test
particle in circular orbit $\Delta A_r$, $\Delta A_r/A_r=2\Delta
v/v$: thus, we have \eqi\rp{\Delta K_p}{K_p}=\rp{\Delta
A_r}{A_r}.\eqf Now, a measurement of the planet's velocity is
needed to get $\Delta K_p/K_p$ (or, equivalently, $\Delta
A_r/A_r$): since $v=na$, where $n$ is the orbital frequency, this
requires a measurement of both $a$ and $n$, while in our case we
only use $a$. Moreover, the measurement of the orbital frequency
pose problems for such planets which have not yet completed a full
orbital revolution, as it was the case for Uranus and Neptune at
the time of the analysis by Anderson et al. (1995). For Neptune,
according to the last row of Table 2 of (Anderson et al. 1995),
$(\Delta K_p/K_p)^{\rm meas}=(-2.0\pm 1.8)\times 10^{-6}$, while
$A_{\rm Pio}/A_{\rm N}=(-133.2\pm 20.3)\times 10^{-6}$. As can be
noted, also in this case the answer is negative  but the accuracy
is far worse than in our test.

\section{Conclusions}
In this paper we used the latest determinations of the secular
extra-rates of the perihelia of some of the  planets of the Solar
System to test two recently proposed gravitational mechanisms to
accommodate the Pioneer anomaly, in the form in which we presently
know it, based on two models of modified gravity. The cleanest
test is for the Brownstein and Moffat (2006) model which, by
fitting a four-free parameters model to the presently available
data from the Pioneer spacecraft, yielded unambiguous predictions
for an extra-acceleration throughout the Solar System. The
determined perihelion rates of Jupiter, Saturn and Uranus neatly
rule out such a nevertheless interesting model. The linear model
originally proposed by Jaekel and Reynaud (2005a; 2005b), along
with its successive non-linear extensions (Jaekel and Reynaud
2006a; 2006b), are, instead, disproved by the determined
perihelion rates both of the inner planets of the Solar System,
especially Mars, and of the outer planets. We also used
short-period effects on the semimajor axis of Uranus and Neptune
and the Voyager 2 ranging data to it to perform another, negative
test.

More generally, in regard to the  impact of a Pioneer-like
extra-acceleration acting upon the celestial bodies lying at the
edge of the region in which the Pioneer anomaly manifested itself
($\sim 20-70$ AU), or entirely residing in it, the present-day
situation can be summarized  as follows
\begin{itemize}
  \item Uranus ($a=19.19$ AU). 3 model-independent tests
\begin{itemize}
  \item Secular advance of perihelion (almost one century of  optical data processed at IAA-RAS): negative
  \item Right ascension/declination residuals (almost one century of optical data processed with the ephemerides of
  IAA-RAS and JPL-NASA):
  negative
  \item Short-period semimajor axis shift (1 ranging measurement at epoch JD=2446455.25 by JPL-NASA): negative
\end{itemize}
  \item Neptune ($a=30.08$ AU). 2 independent tests
\begin{itemize}
  \item Right ascension/declination residuals (about one century or more of optical data processed with the ephemerides of
  IAA-RAS, JPL-NASA and VSOP82):
  negative
  \item Short-period semimajor axis shift (1 ranging measurement at epoch JD=2447763.67 by JPL-NASA): negative
\end{itemize}

  \item Pluto ($a=39.48$ AU). 1 test
\begin{itemize}
  \item Right ascension/declination residuals (almost one century of optical data processed with the ephemerides of
  IAA-RAS; shorter data set analyzed with JPL-NASA ephemerides):
  negative
\end{itemize}

\end{itemize}
In all such tests the determined quantities$-$processed with the
dynamical theories of JPL and IAA independently and without having
the Pioneer anomaly in mind at all$-$were compared to unambiguous
theoretical predictions based on the effects induced by a radial,
constant and uniform acceleration with the same magnitude of that
experienced by Pioneer 10/11, without making any assumptions about
its physical origin.

In conclusion,  it seems to us more and more difficult to
realistically consider the possibility that some modifications of
the current laws of Newton-Einstein gravity may be the cause of
the Pioneer anomaly, at least in its present form, unless a very
strange violation of the weak equivalence principle occurs in the
outer regions of the Solar System (ten Boom 2005). By the way, the
outcome of the currently ongoing re-analysis of the entire
data-set of the Pioneer 10/11 spacecraft  will be of crucial
importance. Indeed, it may turn out that the characteristics of
such an effect  will be different from what we currently know
about it, especially below 20 AU, and/or a satisfactorily
non-gravitational mechanism will finally be found; in any case,
any serious attempt to find a gravitational explanation for the
Pioneer anomaly cannot leave out of consideration the orbital
motions of the Solar System planets.

\section*{Acknowledgements}
I am grateful to E.V. Pitjeva  for her results about the outer
planets of the Solar System and related explanations. Thanks also
to S.Turyshev, O. Bertolami, S. Reynaud, H. Dittus and N. Wright
for stimulating and useful discussions.


\end{document}